\documentclass[aps,pra,noshowpacs,twocolumn,preprintnumbers,superscriptaddress,floatfix]{revtex4}

\usepackage{amsmath,latexsym,amssymb,bm}
\usepackage{hyperref}
\usepackage{color}
\usepackage{graphicx,subfigure}
\usepackage{multirow}
\usepackage{color}
\begin{document}

\title{Transmission of Photonic Polarization States through 55-meter Water:\\
Towards Air-to-sea Quantum Communication}
\author{Cheng-Qiu Hu}
\affiliation{State Key Laboratory of Advanced Optical Communication Systems and
Networks, School of Physics and Astronomy, Shanghai Jiao Tong University, Shanghai 200240, China}
\affiliation{Synergetic Innovation Center of Quantum Information and Quantum Physics,
University of Science and Technology of China, Hefei, Anhui 230026, China}

\author{Zeng-Quan Yan}
\affiliation{State Key Laboratory of Advanced Optical Communication Systems and
Networks, School of Physics and Astronomy, Shanghai Jiao Tong University, Shanghai 200240, China}
\affiliation{Synergetic Innovation Center of Quantum Information and Quantum Physics,
University of Science and Technology of China, Hefei, Anhui 230026, China}

\author{Jun Gao}
\affiliation{State Key Laboratory of Advanced Optical Communication Systems and
Networks, School of Physics and Astronomy, Shanghai Jiao Tong University, Shanghai 200240, China}
\affiliation{Synergetic Innovation Center of Quantum Information and Quantum Physics,
University of Science and Technology of China, Hefei, Anhui 230026, China}

\author{Zhi-Qiang Jiao}
\affiliation{State Key Laboratory of Advanced Optical Communication Systems and
Networks, School of Physics and Astronomy, Shanghai Jiao Tong University, Shanghai 200240, China}
\affiliation{Synergetic Innovation Center of Quantum Information and Quantum Physics,
University of Science and Technology of China, Hefei, Anhui 230026, China}

\author{Zhan-Ming Li}
\affiliation{State Key Laboratory of Advanced Optical Communication Systems and
Networks, School of Physics and Astronomy, Shanghai Jiao Tong University, Shanghai 200240, China}
\affiliation{Synergetic Innovation Center of Quantum Information and Quantum Physics,
University of Science and Technology of China, Hefei, Anhui 230026, China}

\author{Wei-Guan Shen}
\affiliation{State Key Laboratory of Advanced Optical Communication Systems and
Networks, School of Physics and Astronomy, Shanghai Jiao Tong University, Shanghai 200240, China}
\affiliation{Synergetic Innovation Center of Quantum Information and Quantum Physics,
University of Science and Technology of China, Hefei, Anhui 230026, China}

\author{Yuan Chen}
\affiliation{State Key Laboratory of Advanced Optical Communication Systems and
Networks, School of Physics and Astronomy, Shanghai Jiao Tong University, Shanghai 200240, China}
\affiliation{Synergetic Innovation Center of Quantum Information and Quantum Physics,
University of Science and Technology of China, Hefei, Anhui 230026, China}

\author{Ruo-Jing Ren}
\affiliation{State Key Laboratory of Advanced Optical Communication Systems and
Networks, School of Physics and Astronomy, Shanghai Jiao Tong University, Shanghai 200240, China}
\affiliation{Synergetic Innovation Center of Quantum Information and Quantum Physics,
University of Science and Technology of China, Hefei, Anhui 230026, China}

\author{Lu-Feng Qiao}
\affiliation{State Key Laboratory of Advanced Optical Communication Systems and
Networks, School of Physics and Astronomy, Shanghai Jiao Tong University, Shanghai 200240, China}
\affiliation{Synergetic Innovation Center of Quantum Information and Quantum Physics,
University of Science and Technology of China, Hefei, Anhui 230026, China}

\author{Ai-Lin Yang}
\affiliation{State Key Laboratory of Advanced Optical Communication Systems and
Networks, School of Physics and Astronomy, Shanghai Jiao Tong University, Shanghai 200240, China}
\affiliation{Synergetic Innovation Center of Quantum Information and Quantum Physics,
University of Science and Technology of China, Hefei, Anhui 230026, China}

\author{Hao Tang}
\affiliation{State Key Laboratory of Advanced Optical Communication Systems and
Networks, School of Physics and Astronomy, Shanghai Jiao Tong University, Shanghai 200240, China}
\affiliation{Synergetic Innovation Center of Quantum Information and Quantum Physics,
University of Science and Technology of China, Hefei, Anhui 230026, China}

\author{Xian-Min Jin}
\affiliation{State Key Laboratory of Advanced Optical Communication Systems and
Networks, School of Physics and Astronomy, Shanghai Jiao Tong University, Shanghai 200240, China}
\affiliation{Synergetic Innovation Center of Quantum Information and Quantum Physics,
University of Science and Technology of China, Hefei, Anhui 230026, China}

\address{xianmin.jin@sjtu.edu.cn}

\maketitle

\textbf{Quantum communication has been rapidly developed due to its unconditional security and successfully implemented through optical fibers and free-space air in experiment\cite{bennett1984quantum,bouwmeester1997experimental,ekert1991quantum}. To build a complete quantum communication network involving satellites in space and submersibles in ocean, underwater quantum channel has been investigated in both theory and experiment \cite{shi2014feasibility,ji2017towards,bouchard2018underwater}. However, the question of whether the polarization encoded qubit can survive through a long-distance and high-loss underwater channel, which is considered as the restricted area for satellite-borne radio waves, still remains. Here, we experimentally demonstrate the transmission of blue-green photonic polarization states through 55-meter-long water. We prepare six universal quantum states at single photon level and observe their faithful transmission in a large marine test platform. We obtain the complete information of the channel by quantum process tomography. The distance demonstrated in this work reaches a region allowing potential real applications, representing a step further towards air-to-sea quantum communication.}\\

Since the BB84 protocol was proposed by Bennett and Brassard, various protocols and methods have appeared to promote the communication efficiency and increase the secure distance of quantum key distribution (QKD)\cite{bennett1984quantum,bouwmeester1997experimental,ekert1991quantum,lo2005decoy,lo2012measurement,sasaki2014practical,yin2016measurement,lucamarini2018overcoming}. So far, the distance of quantum communication through optical fiber has reached an order of several hundred kilometers\cite{ribordy2000long,tittel2000quantum,jennewein2000quantum,naik2000entangled,dynes2009efficient}, which is applicable for inter-city secure communication\cite{azuma2015all,sun2016quantum,bunandar2018metropolitan}. For the situations without an established fiber link, free-space air has been exploited as a new quantum channel\cite{aspelmeyer2003long,duan2001long,lo2012measurement,jin2010experimental}. From the first 32-cm demonstration in lab to today's quantum satellite, the achieved distance through free-space air has been increased to 1200 km\cite{ren2017ground,yin2017satellite,hwang2003quantum,lo2005decoy,wang2007quantum,huttner1995quantum,brassard2000limitations}.
\begin{figure*}[htb!]
\centering
\includegraphics[width=1.86\columnwidth]{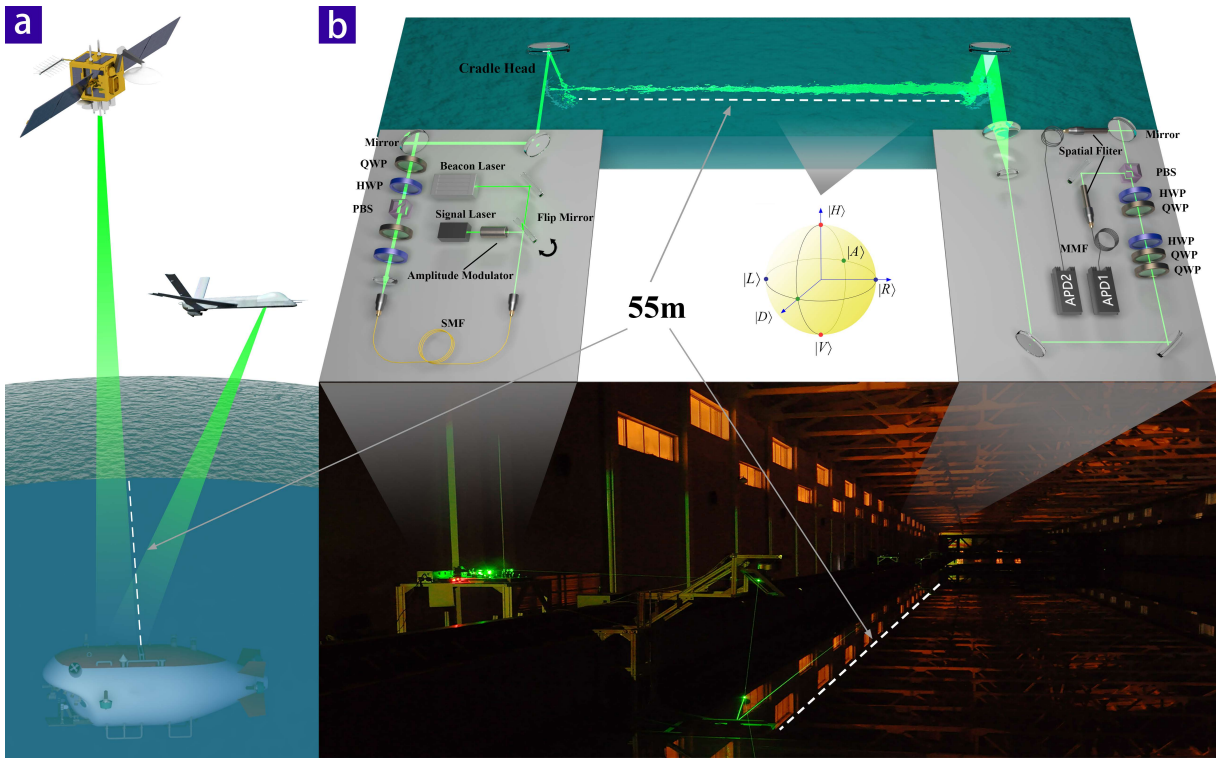}
\renewcommand{\figurename}{Fig.}
\caption{\textbf{Sketch of the experiment.} \textbf{a.} Schematic of the scenarios that have already been practically useful. \textbf{b.} Experimental setup and real field-test environment. The marine test platform is the biggest multiple function towing tank in Asia, with the length of 300m, the width of 16m and the depth of 7.5m. The platform space is actually semi-open in terms of background noises, since there are many windows and they can not be blocked. Six polarization states for testing are marked on Bloch sphere. SMF: single-mode fiber, PBS: polarization beam splitter, HWP: half-wave plate, QWP: quarter-wave plate, MMF: multi-mode fiber.}
\label{fig1}
\end{figure*}

\begin{figure}[b!]
\centering
\includegraphics[width=1\columnwidth]{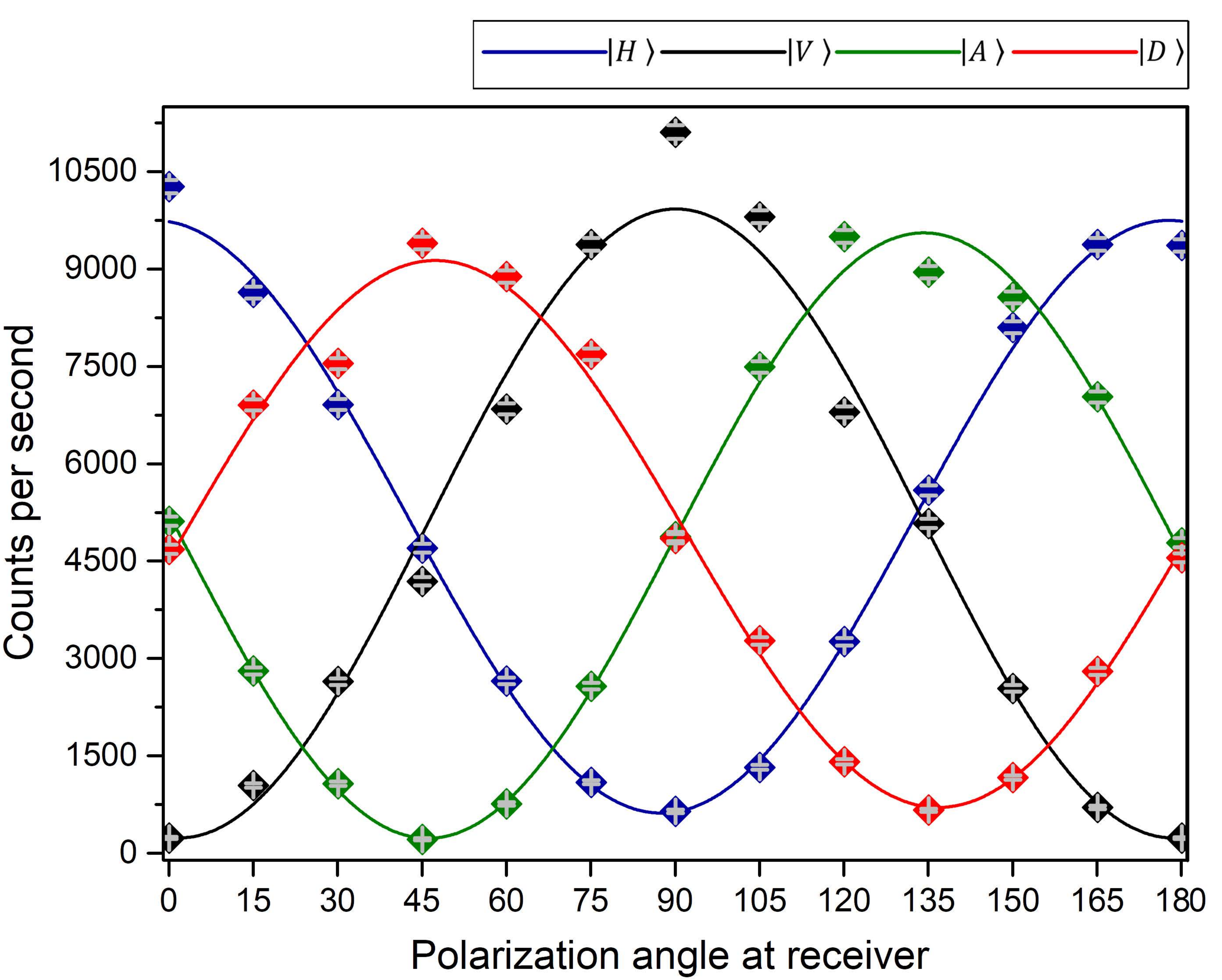}
\renewcommand{\figurename}{Fig.}
\caption{\textbf{Experimental results of polarization correlations between the sent and received states.} Four curves in chart are obtained by projecting the initial states: $|H\rangle$, $|V\rangle$,
$|D\rangle$, $|A\rangle$ at different polarization angles at the receiving terminal. Error bars are too small to be visible.}
\label{fig2}
\end{figure}

Meanwhile, the free-space quantum communication in water environment that covers over 70\% area of the Earth, has attracted much attention for its indispensable role in constructing the global quantum communication network\cite{shi2014feasibility,ji2017towards,bouchard2018underwater}. In the recent experiment in seawater\cite{ji2017towards}, the transmission of polarization-encoded quantum states and quantum entanglement has been demonstrated in a 3.3-meter-long pipe filled with seawater. The obtained high fidelities suggest that underwater quantum communication with photonic polarization is feasible and promising. Very recently, a QKD experiment with photonic orbital angular momentum was performed through a 3-meter underwater channel and the effect of turbulence on transmitted error rates was explored\cite{bouchard2018underwater}.

However, it is still unclear what will happen if single-photon tests go to longer distance and larger tolerance. The theoretically achievable secure communication distances in different single-photon scattering models are not consistent with one another\cite{shi2014feasibility,ji2017towards}. Besides the fundamental interest of verifying theoretical models, it is also very demanding to push the single-photon test into an entirely new region, for boosting applications in special and ultimate scenarios.

For instance, the frequency band of radio wave emitted by communication satellites ranges from 300 MHz to 300 GHz, which can only penetrate the seawater for no more than several meters. Therefore, it is of great importance to experimentally demonstrate an underwater transmission of single-photon states beyond several meters, which may provide a solution of air-to-sea secure communication for submersibles located in open sea.

Here, we experimentally demonstrate the underwater transmission of blue-green photonic polarization states in a large marine test platform up to 55 meters long, as is shown in Fig. 1a, which reaches a distance being able to promise practical quantum communication connecting satellites (or aircraft) and submersibles.

As is shown in Fig. 1b, our experiment is performed in a marine test platform. The platform space is semi-open with many windows connecting the internal and external environment. The sender and the receiver systems locate on the same side of water. Photons from the sender system are steered into water by a wireless-controlled cradle head, and are guided back into free-space air and finally into the receiver system. The whole link consists of two air-water interfaces and a 55-meter-long underwater channel.

The priority in our experiment is to determine the wavelength of photons. There is a blue-green band of 400 nm$\sim$550 nm of which light suffers less attenuation in water\cite{wozniak2007light}. On the one hand, the light at 450 nm possesses the highest transmission rate in pure seawater, but the commercially available silicon avalanche photodiode (APD) is rather inefficient in short wavelength, especially in UV-blue band. On the other hand, the light at longer wavelength has a higher transmission rate in eutrophic water (typically in the Pacific Ocean) than oligotrophic water (typically in the Atlantic Ocean). In view of the trade-off above, we eventually choose the wavelength of 532 nm for our experimental test.

\begin{figure*}[t]
\centering
\par

\includegraphics[width=2.05\columnwidth]{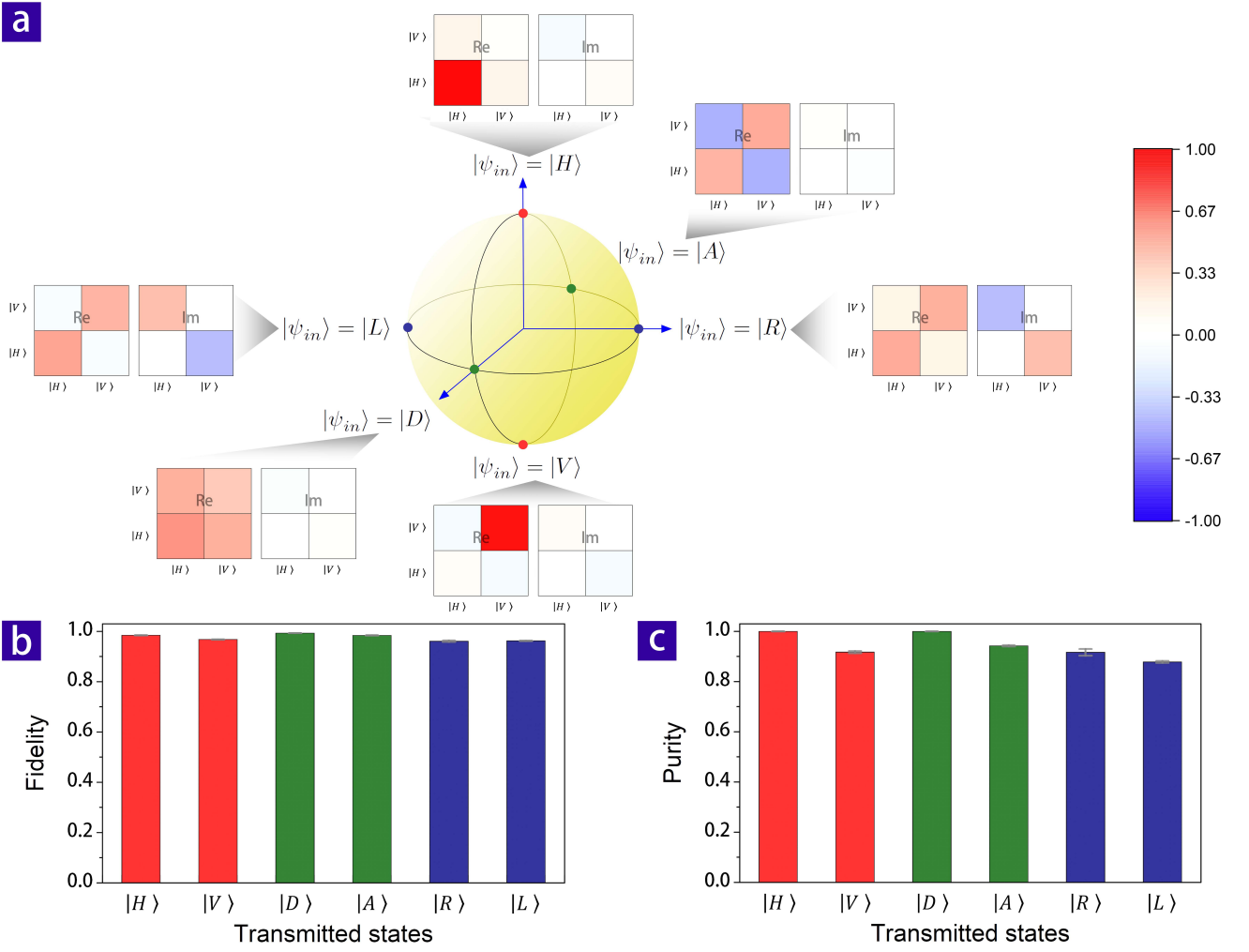}
\par
\renewcommand{\figurename}{Fig.}
\caption{\textbf{Experimental results of underwater transmission of photonic polarization states up to 55 meters.} \textbf{a.} The measured density matrices of the received polarization states are presented together with the states marked on Bloch sphere. \textbf{b.} The fidelities obtained by quantum state tomography. All the fidelities are over 0.95 and the average fidelity is 0.976. \textbf{c.} The purities obtained by quantum state tomography. The average purity is 0.942. The error bars are too small to be visible in the histograms.}
\label{fig3}
\end{figure*}

We employ two laser systems at the sending terminal as the signal and the beacon individually, as shown in Fig. 1b. We use a 532 nm laser with a tunable power range of 0$\sim$100 mW as the source of the signal light. With an amplitude modulator, we prepare the quasi-single photon source with an intensity required in standard decoy-state protocols. The output power is as low as $1.377\times 10^{-19}$ W, which is equivalent to a source with an average number of photons per pulse $\mu =0.37$ at a repetition rate of 1 GHz. In addition, a high power beacon light with a power up to 10 W is used here for the task of underwater alignment. The two lasers are both coupled into a single mode fiber to guarantee a good spatial overlap, meanwhile are switchable with a flip mirror.

There are many difficulties for alignment at 1.5-meter-deep underwater environment. Besides the precise pointing control with cradle head and mirrors, we build a two-lens telescope to expand the light beam to increase its Rayleigh length, for better stability and transmission efficiency. The diameters of the lens are 150 mm ($f$=500 mm) and 25 mm ($f=50$ mm) respectively. The diameter of the collimated beam is about 2 mm at the receiving terminal. We use a combination of HWP  and QWP to compensate polarization rotation induced by the fiber and the other linear optical devices. The received photons are finally projected on a PBS before being coupled into multi-mode fibers.

By comparing the power between the sending terminal and receiving terminal, we are able to estimate the underwater channel loss. The measured attenuation coefficient is about $\alpha=0.16 m^{-1}$, which is close to the coefficient in coastal seawater ($\alpha=0.179 m^{-1}$)\cite{cochenour2008characterization}. The loss of the overall system is about 40 dB. To retrieve signal photons from very strong background noises at high-loss condition is another main challenge for our experiment. By using two long black tubes as spatial filters at the receiving terminals, we obtain an affordable background noise of 600 $s^{-1}$ on APD1 and 200 $s^{-1}$ on APD2. After the whole alignment process, we are ready to switch the beacon light to the signal light.

We prepare six universal polarization-encoded quantum states at the sending terminal with the combination of a polarization beam splitter (PBS), a half-wave plate (HWP) and a quarter-wave plate (QWP): $%
\left\vert H\right\rangle ,\left\vert V\right\rangle ,\left\vert
D\right\rangle =\frac{1}{\sqrt{2}}(\left\vert H\right\rangle +\left\vert
V\right\rangle ),\left\vert A\right\rangle =\frac{1}{\sqrt{2}}(\left\vert
H\right\rangle -\left\vert V\right\rangle ),\left\vert R\right\rangle =\frac{%
1}{\sqrt{2}}(\left\vert H\right\rangle +i\left\vert V\right\rangle
),\left\vert L\right\rangle =\frac{1}{\sqrt{2}}(\left\vert H\right\rangle
-i\left\vert V\right\rangle )$. The instability of water causes beam wandering and profile distortion, which both lead to the fluctuation of the photon counting numbers. The received signal intensity is about 11000 $s^{-1}$ in average, fluctuating from 9000 $s^{-1}$ to 12000 $s^{-1}$. Here we use two detectors to detect transmitted and reflected photons under orthogonal projective measurements simultaneously to eliminate the influence of such fluctuation.

\begin{figure}[htb!]
\centering
\includegraphics[width=1\columnwidth]{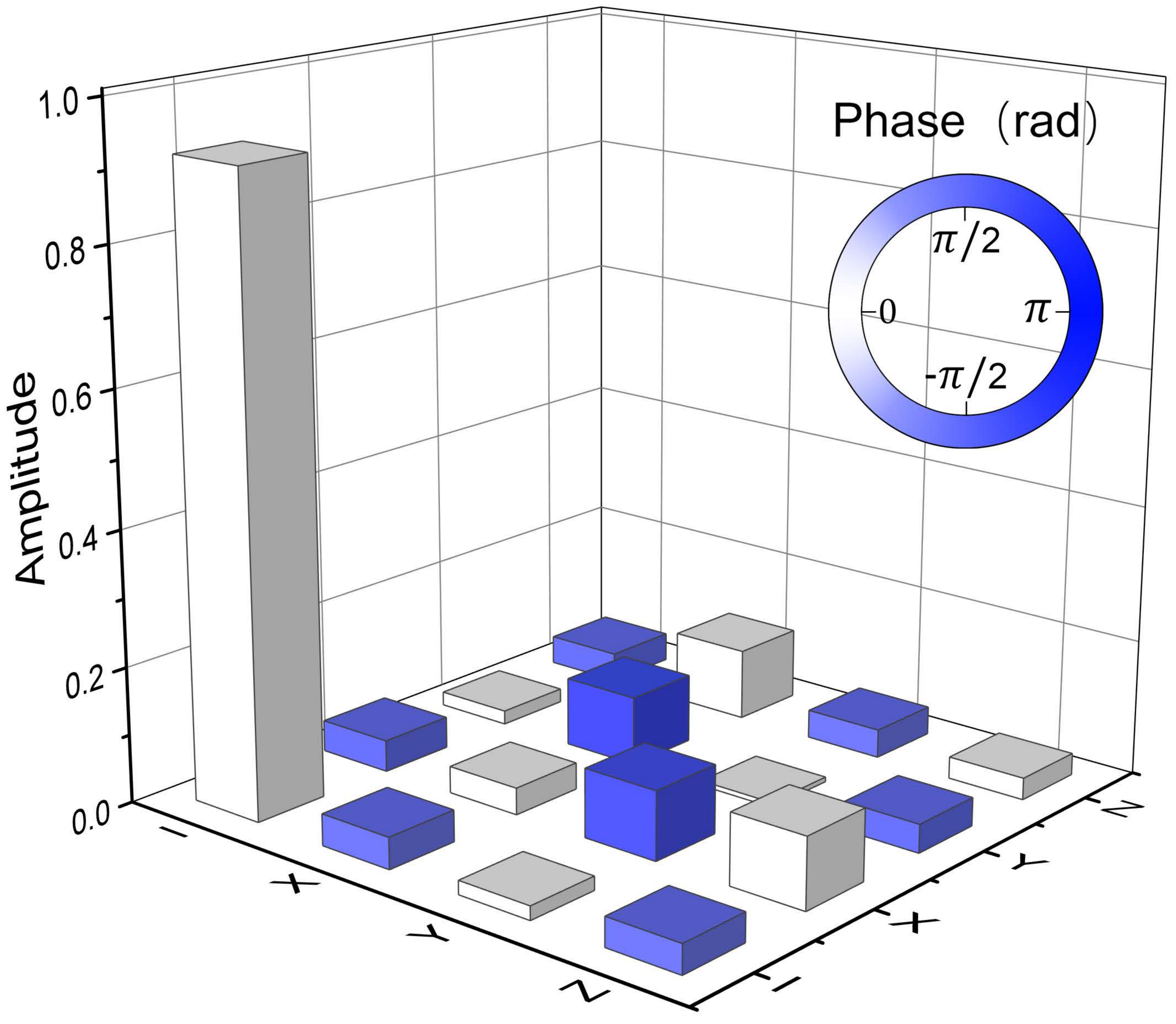}\newline
\renewcommand{\figurename}{Fig.}
\caption{\textbf{Experimental results of quantum process tomography for the 55-meter-long underwater channel} The measured $\chi_{\text{water}}$ matrix in bases of $%
\widetilde{E}_0=I,\widetilde{E}_1=X,\widetilde{E}_2=Y,\widetilde{E}_3=Z$. The modulus and argument of the process matrix elements are represented by the height and color of the bars respectively. The fidelity between the matrix $\chi_{\text{water}}$ and $\chi_{\text{ideal}}$ is 0.96.}
\label{fig4}
\end{figure}

Firstly, we verify the polarization correlations between the sent and received states. As is shown in Fig. 2, we prepare the states at $%
\left\vert H\right\rangle ,\left\vert V\right\rangle ,\left\vert
D\right\rangle ,\left\vert A\right\rangle $ and project them under different polarization angles at the receiving terminal. The visibilities obtained by fitting the sinusoidal curves are $\left\vert H\right\rangle :87\%, \left\vert V\right\rangle
:95.2\%, \left\vert D\right\rangle :85.5\%, \left\vert A\right\rangle :95.2\%$%
. The visibilities in basis H/V and D/A are slightly different due to the unbalanced
background noises on the two detectors. The average visibility is $90.7\%$ without subtracting the background noises.

To obtain comprehensive and complete information of the received states, we implement quantum state tomography\cite{PhysRevA.64.052312}. The density matrices of the six universal states are shown in Fig. 3a. From the density matrices we calculate the state fidelity according to $F_{s}= \text{tr} \sqrt{\sqrt{\rho _{\text{received}}}\rho _{\text{%
sent}}\sqrt{\rho _{\text{received}}}}$\cite{nielsen2010quantum,JOHANSSON20121760} and present the results in Fig. 3b. We can see that all fidelities are over 95\% and the average fidelity is up to 0.976. We further calculate the purity of all these states. As plotted in Fig. 3c, the average purity of the states is up to 0.942.

In order to further reveal the physical process of the underwater channel, we also perform quantum process tomography. Here we use the matrix $\chi $ to represent the effect of the underwater channel, and the output state can be written as
\begin{equation}
\varepsilon (\rho
)=\sum\limits_{mn}\widetilde{E}_{m}\rho \widetilde{E}_{n}^{\dag }\chi _{mn}.
\end{equation}
As for one-qubit process tomography, we choose $\widetilde{E}_{m}$ as: $\widetilde{E}_{0}=I$, $\widetilde{E}_{1}=X$, $\widetilde{E}_{2}=Y$, $\widetilde{E}_{3}=Z$. The measured results of quantum process tomography are shown in Fig. 4. Regardless of the loss effect, the operation conducted by the underwater channel is found similar to the ideal case:
\begin{equation}
\chi _{\text{ideal}}=%
\begin{bmatrix}
1 & 0 & 0 & 0 \\
0 & 0 & 0 & 0 \\
0 & 0 & 0 & 0 \\
0 & 0 & 0 & 0%
\end{bmatrix}%
\end{equation}
which is the cornerstone for being a reliable quantum channel. We can also calculate the fidelity of the process matrix according to the formula $F_{P}=\text{tr}\sqrt{\sqrt{\chi _{\text{water}}}\chi _{\text{ideal}}\sqrt{\chi _{\text{water}}}}
$, and we obtained a result of up to 0.96, which suggests that photonic polarization states can well survive under the conditions of strong scattering and high loss through long underwater channel.

In summary, we experimentally demonstrate the transmission of blue-green photonic polarization states through a 55-meter-long underwater channel. The obtained state and process fidelities confirm the feasibility of implementing secure quantum communication with submersibles located in open sea. It should be noticed that the attenuation coefficient in our marine test platform only approaches the quality of costal sea.
The results in blue-green window for open sea have been found as low as 0.018 $m^{-1}$  \cite{jerlov1976marine,cochenour2008characterization}, which suggests an even deeper achievable distance. In addition, as the blue-green band is a window of light for both seawater and atmosphere, we can therefore expect a long-distance air-to-sea quantum communication from satellites to submersibles in a depth forbidden to the radio wave frequency in the future.

\section*{Acknowledgements} The authors thank Hang Li, Xiao-Ling Pang, Jian-Peng Dou and Jian-Wei Pan for helpful discussions. This work was supported by National Key R\&D Program of China (2017YFA0303700); National Natural Science Foundation of China (NSFC) (61734005, 11761141014, 11690033); Science and Technology Commission of Shanghai Municipality (STCSM) (15QA1402200, 16JC1400405, 17JC1400403); Shanghai Municipal Education Commission (SMEC)(16SG09, 2017-01-07-00-02-E00049); X.-M.J. acknowledges support from the National Young 1000 Talents Plan.

\bibliographystyle{apsrev4-1}
%

\end{document}